\documentclass{ws-p8-50x6-00}

\begin{document}

\title{Elliptic flow from an on-shell parton cascade}

\author{D. Moln\'ar \lowercase{AND} M. Gyulassy}

\address{Physics Department, Columbia University\\ 538 West 120th Street, New York, NY 10027, U.S.A. \\
E-mail: molnard@phys.columbia.edu, gyulassy@nt3.phys.columbia.edu}


\maketitle

\abstracts{
Differential elliptic flow out to $p_\perp\sim 5$ GeV/c
is calculated using the MPC elastic parton
cascade model\cite{MPC,nonequil} for Au+Au at
$E_{cm}\sim 130$ $A$ GeV.
The results are compared to recent
STAR/RHIC elliptic flow data\cite{STARv2}. 
An elliptic flow pattern comparable to the data seems to require
initial parton densities at least twice higher
than the predicted by the HIJING model\cite{Gyulassy:1994ew}.
Elliptic flow is also shown to be sensitive to the hadronization procedure.
}

\section{Introduction} 
\label{Section:intro}

Elliptic flow, $v_2(p_\perp)=\langle \cos(2\phi)\rangle_{p_\perp}$,
 the differential second moment of the azimuthal momentum distribution,
has been the subject of increasing interest
\cite{hydro,Zhang:1999rs,molnar:QM99,Wang:2000fq,Gyulassy:2000gk}
since the discovery\cite{STARv2} at RHIC that $v_2(p_\perp>1 $ GeV)
 $\rightarrow 0.2$.
This sizable  high $p_\perp$ collective effect
 depends strongly on the dynamics in a heavy
ion collision and provides important information about the 
density and effective energy loss of partons.

The simplest theoretical framework to study elliptic flow is ideal
hydrodynamics\cite{hydro}.
For RHIC energies, ideal hydrodynamics
agrees remarkably well with the measured elliptic flow data%
\cite{STARv2}
up to transverse momenta $\sim 1.5$
GeV$/c$.
However,
it fails to saturate at high $p_\perp>2$ GeV as does the data reported
by STAR at Quark Matter 2001.

A theoretical problem with ideal hydrodynamics is that it
assumes local equilibrium throughout the whole evolution.
This idealization is marginal
for conditions encountered in heavy ion collisions\cite{nonequil}.
A theoretical framework is required that allows for nonequilibrium dynamics.
Covariant  Boltzmann transport theory provides a convenient framework
that depends on the local mean free path $\lambda(x) \equiv 1/\sigma n(x)$.
It interpolates between  free streaming ($\lambda=\infty$)
and  ideal hydrodynamics ($\lambda=0$).

Parton cascade simulations\cite{Zhang:1999rs,molnar:QM99,v2}
show on the other hand, that  the initial parton density
based on HIJING\cite{Gyulassy:1994ew}
is too low  to produce the observed  elliptic flow
unless the pQCD cross sections are artificially enhanced 
by a factor $\sim 2-3$.
However,
gluon saturation models\cite{Eskola:2000fc}
predict up to five times higher initial densities,
and these  may be dense enough to generate the observed collective flow
even with pQCD elastic cross sections.
In this study,
we explore the dependence of elliptic flow on the initial density,
or equivalently%
\footnote{
The equivalence is due to the scaling property explained in Section
\ref{Subsection:subdivision}.},
on the partonic cross section.

Calculations
based on inelastic parton energy loss\cite{Wang:2000fq,Gyulassy:2000gk}
also predict saturation or  decreasing $v_2$ at high $p_\perp$.
These calculations are only valid for high $p_\perp$,
where collective transverse flow from lower $p_\perp$ partons can be neglected.
The hydrodynamic component from low $p_\perp$
is, 
on the other hand,
automatically incorporated in parton cascades.
Though parton cascades
lack at present covariant
inelastic energy loss, 
we find 
in Fig. \ref{Figure:v2}
that elastic energy loss alone may  account
for the observed high $p_\perp$ azimuthal flow pattern
as long as the elastic opacity is 
large enough.

We compute the partonic evolution with MPC\cite{MPC},
a newly formulated, covariant, parton kinetic theory technique.
MPC is 
an extension of the covariant parton cascade algorithm,
ZPC\cite{Zhang:1998ej}.
Both MPC and ZPC have been
extensively tested\cite{Gyulassy:1997ib,Zhang:1998tj} 
and compared to analytic transport solutions
and covariant Euler and Navier-Stokes dynamics in 1+1D geometry.
A critical new element of both these algorithms
is the parton subdivision technique proposed by Pang\cite{Zhang:1998tj,Yang}.

\section{Covariant Parton Transport Theory}
\label{Section:transport_theory}

We consider here, as in  Refs.\cite{MPC,nonequil,Zhang:1998ej,Yang},
the simplest form of Lorentz-covariant Boltzmann transport theory
in which the on-shell phase space density $f(x,\vp)$,
evolves with an elastic $2\to 2$ rate as
\bea
p_1^\mu \partial_\mu f_1 &=&\int\limits_2\!\!\!\!
\int\limits_3\!\!\!\!
\int\limits_4\!\!
\left(
f_3 f_4 - f_1 f_2
\right)
W_{12\to 34} \delta^4(p_1+p_2-p_3-p_4)
+  \, S(x, \vp_1) .
\label{Eq:Boltzmann_22}
\eea
Here $W$ is the square of the scattering matrix element,
the integrals are shorthands
for $\int\limits_i \equiv \int \frac{g\ d^3 p_i}{(2\pi)^3 E_i}$,
where $g$ is the number of internal degrees of freedom,
while $f_j \equiv f(x, \vp_j)$,
and we  interpret  $f(x,\vp)$ as describing
an ultrarelativistic massless gluon gas 
with $g=16$ (8 colors, 2 helicities).
The initial conditions are specified by the source function $S(x,\vp)$,
which we discuss in Section \ref{Section:results}.

The elastic gluon scattering matrix elements in dense parton systems
are modeled by a Debye-screened form
\be
\frac{d\sigma}{dt}
  = \sigma_0
    \left(1 + \frac{\mu^2}{s}\right)
    \frac{\mu^2}{\left(t-\mu^2\right)^2},
\label{Eq:cross_section}
\ee
where $\mu$ is the screening mass,
$\sigma_0 = 9\pi\alpha_s^2/2\mu^2$ is the total cross section,
which we chose to be independent of energy.
For a fixed $\sigma_0$,
the relevant transport cross section
$\sigma_t = \int d\sigma \sin^2\theta_{cm}$
is maximal in the isotropic ($\mu\to\infty$) case.

It is important to emphasize that while the cross section suggests a
geometrical picture of action over finite distances, 
Eq. (\ref{Eq:cross_section}) is only a convenient parametrization to
describe the effective {\em local} transition probability, $W$,
as $dW/dt=s\;d\sigma/dt$.
The particle subdivision technique
(see next Section) removes all notion of
nonlocality in this approach, as  in hydrodynamics.

\section{Parton Subdivision and Scaling of Solutions}
\label{Subsection:subdivision}

Cascade algorithms inevitably 
violate Lorentz covariance
because particle interactions are assumed to occur
whenever the distance of closest approach (in the relative c.m.)
is $d<\sqrt{\sigma_0/\pi}$, which corresponds to action at a distance.
To recover the {\em local} character of equation (\ref{Eq:Boltzmann_22})
and hence Lorentz covariance,
we use the parton subdivision technique\cite{Zhang:1998ej,Yang}.
This technique is based  on the covariance of Eq. (\ref{Eq:Boltzmann_22})
under the transformation 
\be f\to f'\equiv l\, f, \quad W\to W'\equiv W/l \quad
(\sigma\to \sigma' = \sigma/ l),
\label{Eq:particle_subdivision}
\ee
which effectively rescales the cascade interaction range by $1/\sqrt{l}$.
Lorentz violation formally vanishes\cite{Zhang:1998tj}
in the $l\to \infty$ limit.
In practice, very high $\sim 100-1000$ subdivisions
are needed\cite{nonequil} to obtain accurate numerical solutions
from initial conditions expected at RHIC.

Subdivision covariance (\ref{Eq:particle_subdivision})
implies that the transport equation has a broad dynamical range,
and the solution for any given initial condition and transport property
immediately provides the solution to a broad band
of suitably scaled initial conditions and transport properties. 
This is because solutions
for problems with $l$ times the initial density
$dN/d\eta$,
but with one $l$-th the reaction rate
can be mapped to the original ($l=1$) case. 
We must use subdivision to eliminate numerical artifacts.
However, once that is achieved,
we have actually found the solution to a whole class of 
suitably rescaled problems. 

The dynamical range of Eq. (\ref{Eq:Boltzmann_22}) 
is further increased by its covariance under coordinate 
and momentum rescaling\cite{nonequil},
leading to covariance under
\bea
f(x,\vec p) &\to& f'(x,\vec p)
\equiv l_p^{-3} l \,f\!\left(\frac{x}{l_x},\frac{\vec p}{l_p}\right),
\nonumber\\
W(\{p_i\}) &\to& W'(\{p_i\})
\equiv \frac{l_p^2}{l_x l}\, W\!\left(\left\{\frac{p_i}{l_p}\right\}\right),
\quad m \to  m'=m/l_p,
\label{Eq:rescale_all}
\eea
where $l_x$ and $l_p$ are the coordinate and momentum scaling parameters,
respectively.
This means\cite{nonequil}
that we can scale one solution to others 
provided that
$\mu/T_0$, $\sigma_0 dN/d\eta$ and $R_0 / \tau_0$ remain the same.

\section{Results and Discussion}
\label{Section:results}


\begin{figure}[hptb]
\vspace*{-1cm}
\center
\leavevmode
\hspace*{-0.6cm}
\hbox{
    \epsfysize 5.5cm
    \epsfbox{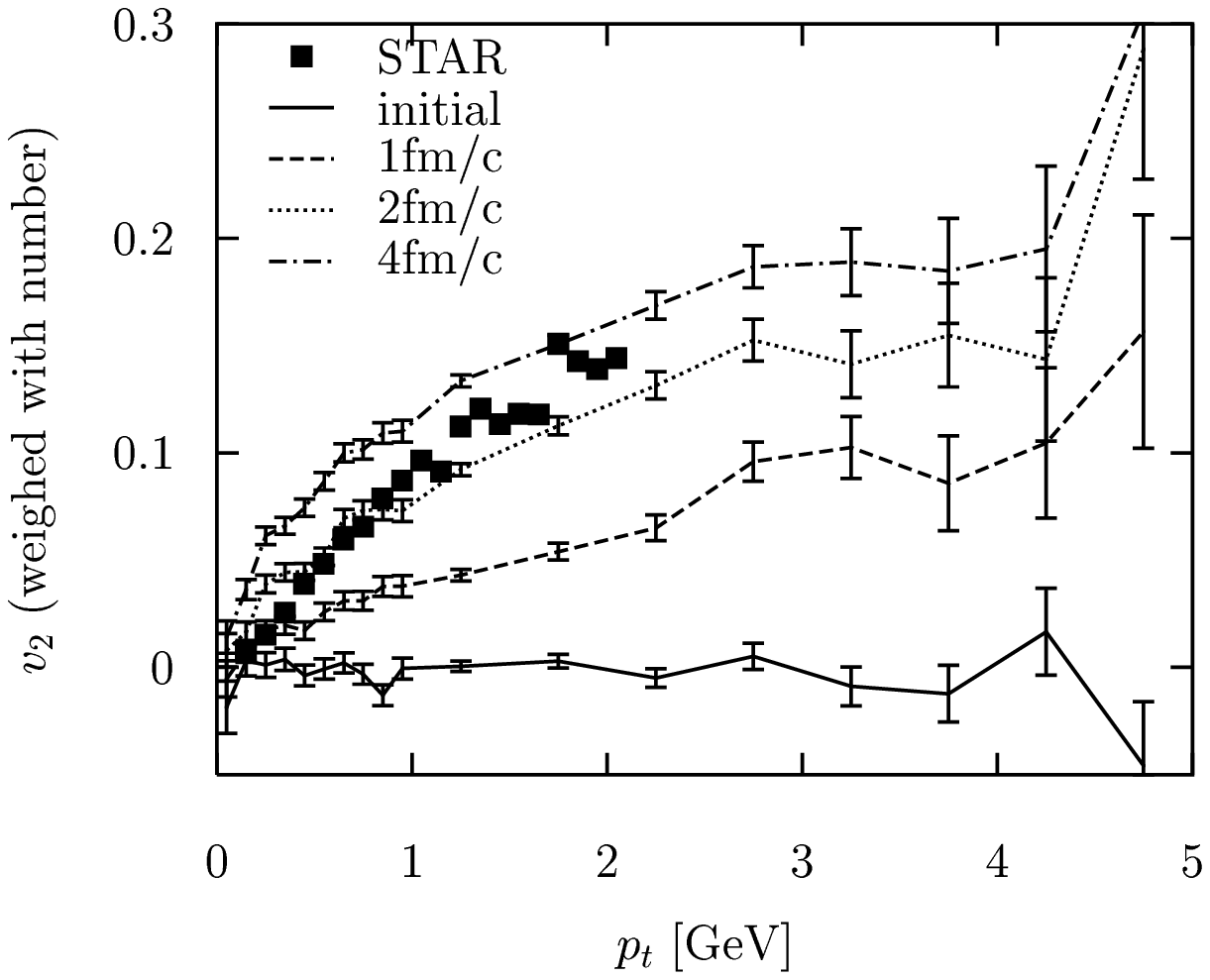}
\hskip -0.5cm
    \epsfysize 5.5cm
    \epsfbox{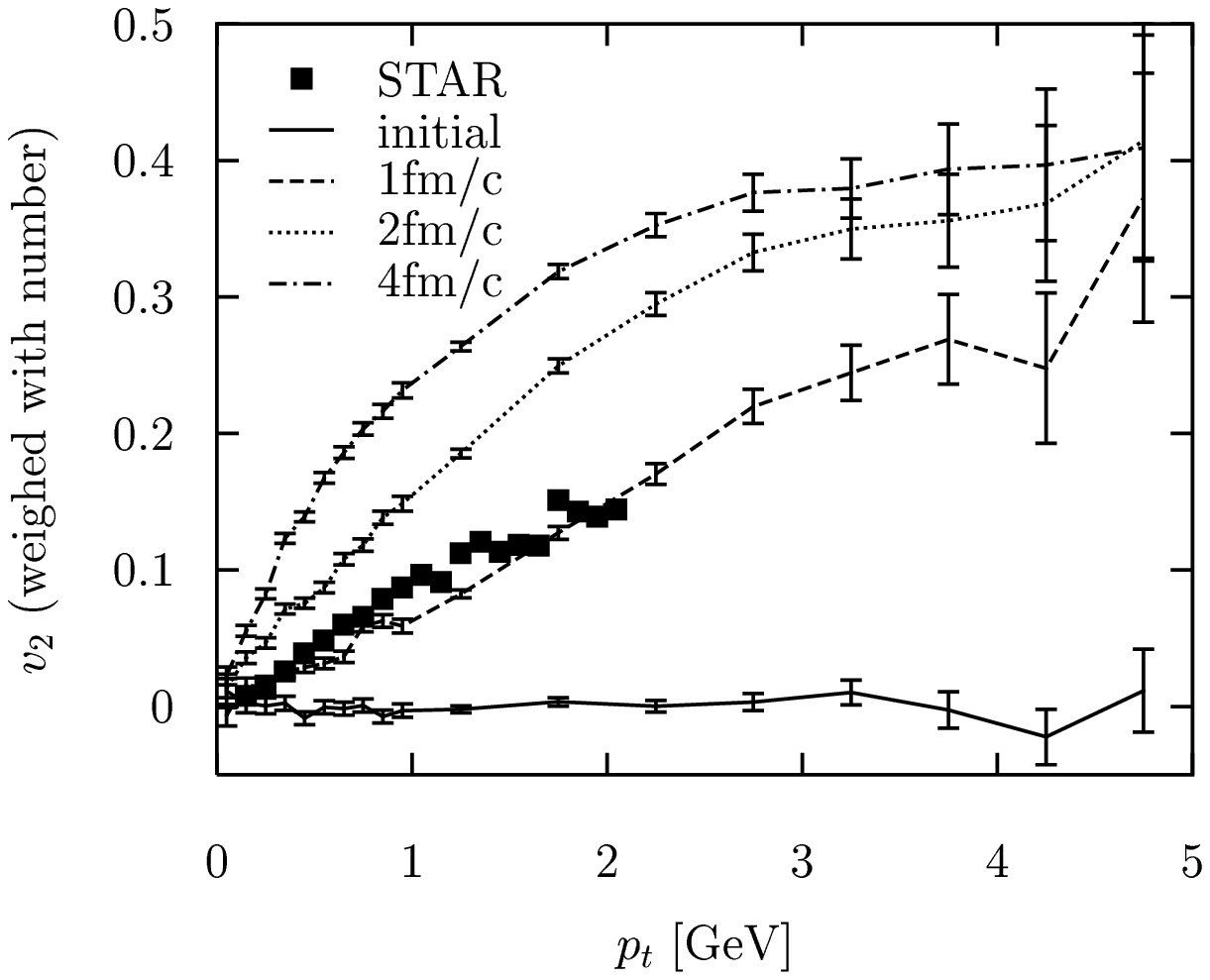}
}

\vspace*{-0.7cm}
\hspace*{-0.6cm}
\hbox{
    \epsfysize 5.5cm
    \epsfbox{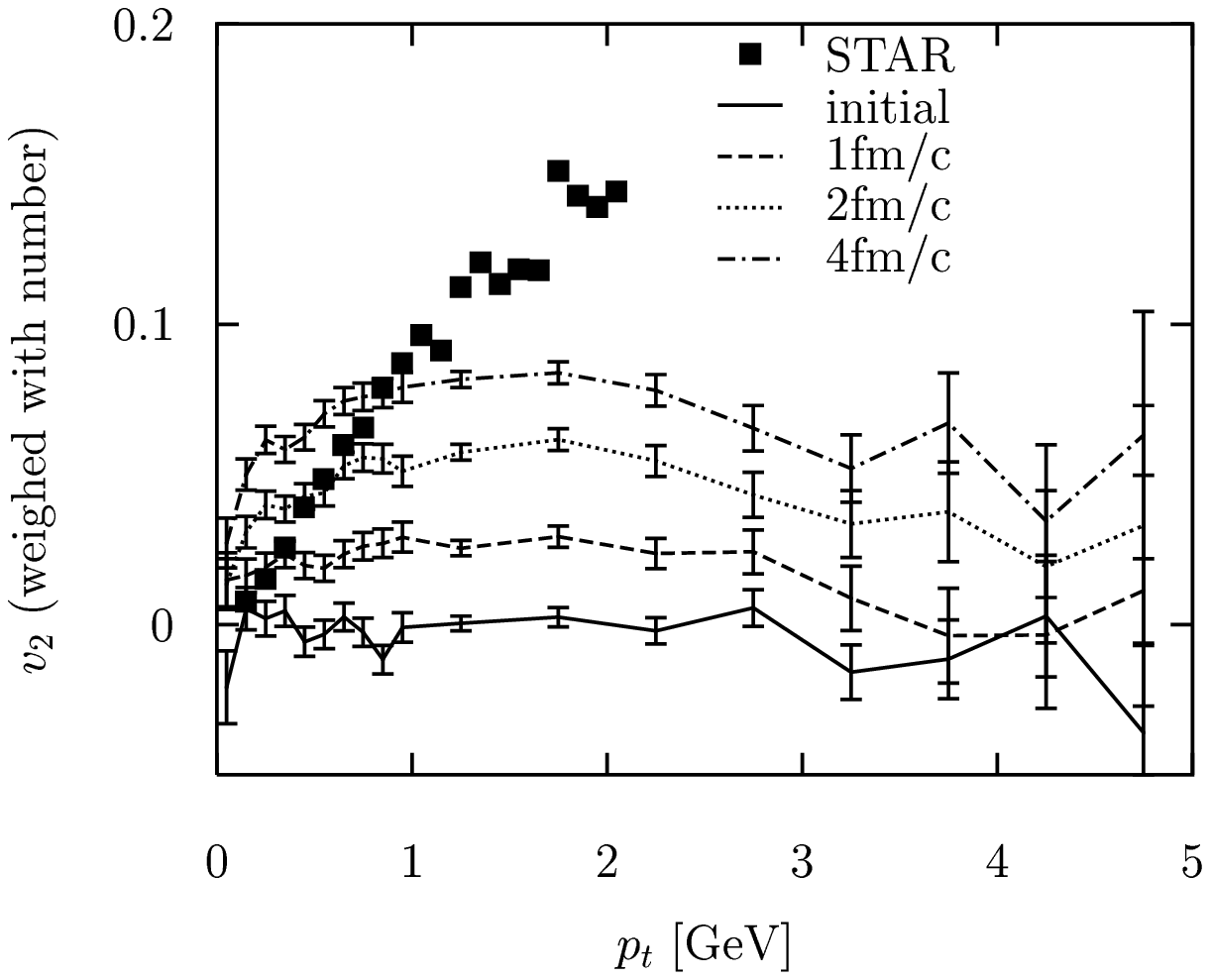}
\hskip -0.5cm
    \epsfysize 5.5cm
    \epsfbox{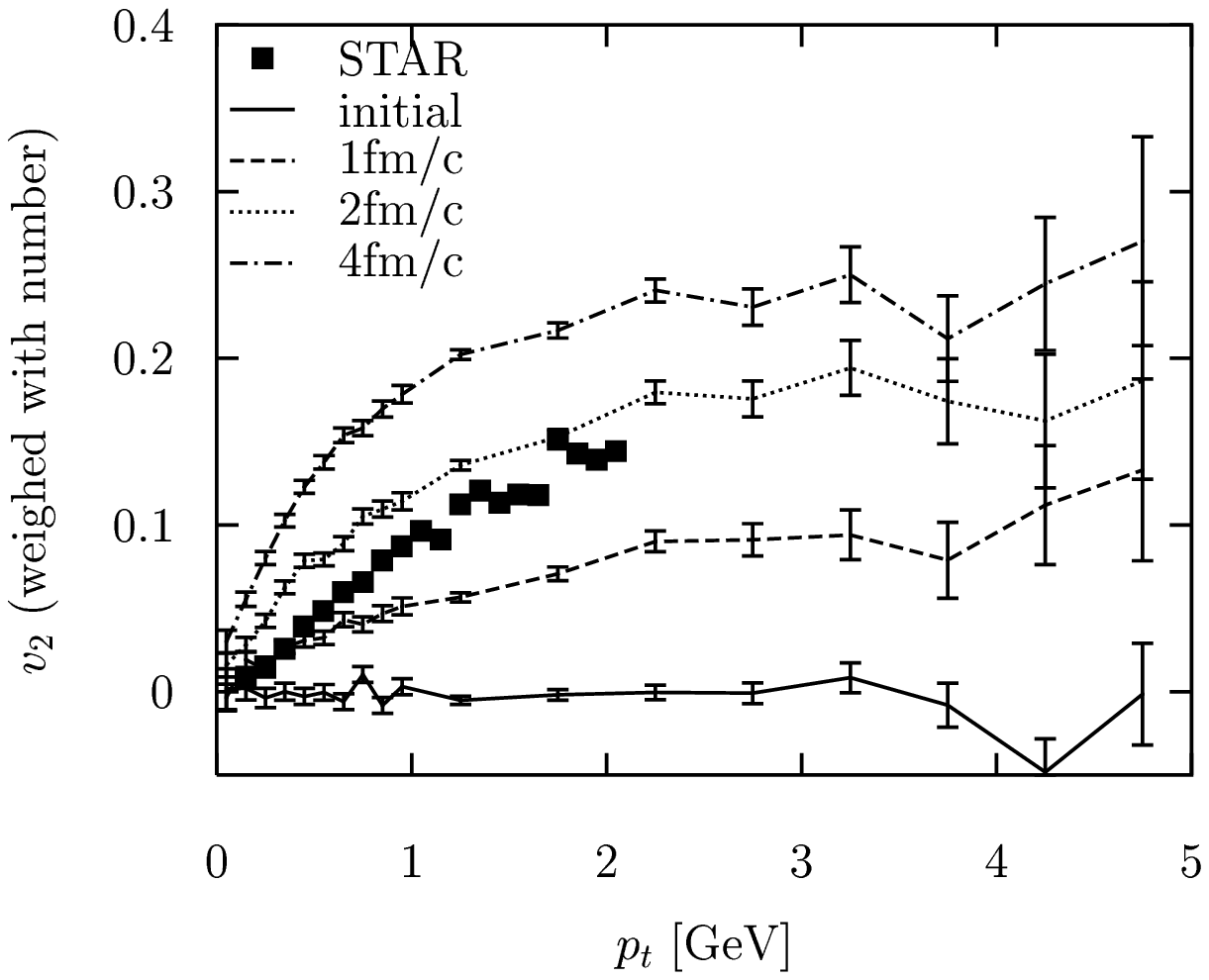}
}
\vspace*{-0.7cm}
\caption{
\footnotesize
The proper time evolution of elliptic flow as a function of transverse momentum,
for an Au+Au collision at $\sqrt{s}=130A$ GeV with $b=8$ fm,
in case of isotropic (top row) and gluonic (bottom row)
cross section with $\sigma_0 = 3$ mb (left) and 10 mb (right).
}
\label{Figure:1}
\label{Figure:v2}
\vspace*{-0.6cm}
\end{figure}

Following Ref.\cite{Zhang:1999rs}, we modeled
the colliding nuclei as
longitudinally boost invariant Bjorken cylinders
with radii $R_0 = 6$ fm
in local thermal equilibrium
at temperature $T_0$ at proper time $\tau_0=0.1$~fm/$c$.
The  pseudo-rapidity $\eta\equiv 1/2 \log((t+z)/(t-z))$ distribution
was taken as uniform between $|\eta| < 5$.
For central collisions, $T_0 = 500$ MeV and $dN/d\eta = 400$
as by fitting the gluon mini-jet transverse momentum spectrum
predicted by HIJING\cite{Gyulassy:1994ew}
(including shadowing and jet quenching effects).
For non-central collisions,
as a crude estimate, we scaled $dN/d\eta$ by the overlap area of 
the two cylinders.
To maximize elliptic flow,
the impact parameter was chosen to be $b = 8$ fm.

Evolutions from different initial densities
can be obtained by varying the cross section only
and using the scaling property explained in Section \ref{Subsection:subdivision}.
Our simulations were performed both with isotropic and gluonic cross sections
($\mu = 0.5$ GeV)
with $\sigma_0=3$ and 10 mb.
The particle subdivision was $l=100$.

Fig. \ref{Figure:v2}
shows the proper time evolution of the gluon elliptic flow as a function
of transverse momentum for the different cross sections.
For all cross sections studied,
elliptic flow reaches its final asymptotic value
early, by $\tau = 4$ fm$/c$,
reinforcing results in Ref. \cite{Zhang:1999rs}.
Also, elliptic flow increases with increasing {\em transport} cross section,
as expected.

Elliptic flow monotonically increases until
$p_\perp \sim 2$ GeV, where it saturates.
According to a more detailed study\cite{v2},
this saturation occurs for all impact parameters,
leading to a saturation in the impact parameter averaged (minimum bias)
elliptic flow as well.
For the 3 mb gluonic cross section,
$v_2$ decreases after 2 GeV,
 due to elastic energy loss.


\begin{figure}[hptb]
\vspace*{-0.7cm}
\center
\leavevmode

    \epsfysize 5.5cm
    \epsfbox{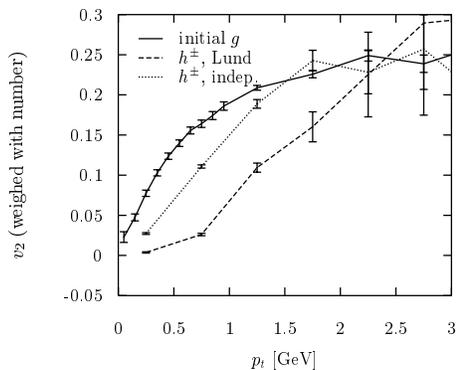}

\vspace*{-0.5cm}
\caption{
\footnotesize
Partonic and the final hadronic elliptic flow after hadronization
via independent or Lund fragmentation,
for the 10 mb gluonic cross section.
}
\label{Figure:2}
\label{Figure:v2_hadronization}
\vspace*{-0.4cm}
\end{figure}

Fig. \ref{Figure:v2_hadronization} shows the effect of hadronization
on elliptic flow.
Two extreme cases are presented:
independent fragmentation
and Lund fragmentation via 400 long
($\Delta y = 10$) strings.
In the case of independent fragmentation,
the produced hadrons move in slightly different direction
than the original parton (jet cone effect),
resulting in a moderate reduction of elliptic flow.
Lund fragmentation produces all hadrons with almost complete
azimuthal symmetry
\footnote{
The azimuthal symmetry is not perfect because the gluons kink the strings,
however, the kinks are very small.
},
leading to an order of magnitude reduction at low $p_\perp$. 

\section{Conclusion}
The elliptic flow results from the parton
cascade MPC\cite{MPC,nonequil}
strongly suggest that the observed saturation of elliptic flow 
at $p_\perp \sim$ 2 GeV can be {\em quantitatively} reproduced
from parton kinetic theory,
if the initial parton density is high enough.
In addition, the sensitivity of elliptic flow on hadronization scheme
will provide a strong constraint on possible hadronization scenarios.
A more detailed study of these issues is in progress\cite{v2}.

\vspace*{-0.1cm}
\section*{Acknowledgments}
We acknowledge the Parallel Distributed Systems Facility
at the National Energy Research Scientific Computing Center
for providing computing resources.

This work was supported by the Director, Office of Energy Research,
Division of Nuclear Physics of the Office of High Energy and Nuclear Physics
of the U.S. Department of Energy under contract No. DE-FG-02-93ER-40764.

\end{document}